\documentclass[prd,aps, nofootinbib, byrevtex,showpacs,
preprint,
amsmath,amssymb]{revtex4}
\usepackage{bm}
\usepackage{graphics}
\usepackage{float}
\usepackage{latexsym}
\usepackage{epsfig}
\usepackage{dcolumn}
\usepackage{url}
\usepackage{lmodern} %math, rm, ss, tt
\usepackage[T1]{fontenc}

\usepackage[pdftex,bookmarks,colorlinks,breaklinks]{hyperref}
\hypersetup{linkcolor=black
,citecolor=black %blue
,filecolor=black} %dullmagenta
\usepackage{epstopdf}% to include .eps graphics files with pdfLaTeX
\begin{document}

\title{Forms of matter and forms of radiation}
\vspace{5cm}
 \author{Maurice Kleman}\email{maurice.kleman@mines.org}
 %\vspace{0.5cm}
\affiliation{\vspace{1cm}Institut de Physique du Globe de Paris\\
4, place Jussieu - 75252 Paris cédex 05\\
\noindent(UMR CNRS 7154)}%\vspace{0.5cm}
\begin{abstract}
\vspace{0.5cm}
The theory of defects in ordered and ill-ordered media is a well-advanced part of condensed matter physics.  
Concepts developed in this field also occur in the study of spacetime singularities, namely: i)-  the {topological theory of quantized defects} (Kibble's cosmic strings) 
and ii)- the {Volterra process for continuous defects}, used to classify the Poincar\'e symmetry breakings. We reassess the classification of Minkowski spacetime defects in the same theoretical frame, starting from
the conjecture that these defects fall into two classes, as on they relate to massive particles or to radiation. This we justify on the empirical evidence
of the Hubble's expansion. We introduce timelike and null congruences of geodesics treated as {ordered media}, viz. \textsf{$m$-crystals} of massive particles and \textsf{$r$-crystals} of massless particles, with parallel 4-momenta in $M^4$. Classifying their defects (or \textsf{forms}) we find (i)
$m$- and $r$- {Volterra continuous line defects} and (ii) {quantized topologically stable $r$-defects}, these latter forms being of various dimensionalities. Besides these \textsf{perfect} forms, there are \textsf{imperfect disclinations} that bound misorientation walls in three dimensions.
We also speculate on the possible relation of these forms with the large-scale structure of the Universe.

%199 words

\end{abstract}

\date{March 23, 2011}

\maketitle

%\newpage
%\scriptsize
%\tableofcontents
%\vspace{2cm}
\newpage
\normalsize
\section{Introduction}
Cosmology is a science in explosion; the number of new and imaginative models of the History of the Universe (or the Universes), in concomitance with the publication of a considerable number of books of popularization, outpasses the possibilities of the serious reader. 
The vast majority of these works go by far beyond the concepts elaborated in the frame of General Relativity (GR), as if these concepts have already borne fruit and as if further taking them in consideration is useless. But is it really the case? 
Concepts belonging to the theory of spacetime defects are e.g. those of Hawking \& Penrose (geodesic singularities  \cite{penrose65,hawk73}), and Kibble (cosmic strings \cite{kibble76,vilenkin94}).  
Both approaches refer to spacetime defects of completely different kinds. A third type of spacetime singularities has also been introduced (only in flat Minkowski spacetime) that are closer in spirit to the usual defects of Condensed Matter Physics (CMP), see e.g. \textcite{ruggiero03} for a review. 
This paper
proposes a new approach inspired by condensed matter defect theory and entirely in the frame of classical gravitation theory; it introduces defects specific of regions of spacetime filled homogeneously either with a cloud of dust or with a beam of radiation propagating in spacetime with a homogeneous constant 4-momentum. 
We distinguish
 \begin{enumerate}
\item 
\textsf{forms of matter} ({$m$-forms}): these are the defects of various dimensionalities carried by a cloud of dust. 
These defects are not topologically stable. Those which are 2-dimensional in 4D spacetime (1-dimensional in 3D space) are akin to the Volterra continuous defects of CMP,
  \item
\textsf{forms of radiation} ({$r$-forms}): these defects, carried by a beam of massless particles e.g. photons, are of three types: (i)- \textit{line defects} in  spacelike slices (2-dimensional defects in 4D spacetime), which are not topologically stable, (ii)- topologically stable \textit{point defects} (1-dimensional defects in 4D spacetime), (iii)- topologically stable \textit{singular events}, 
  \item 
\textsf{boost disclinations}: these defects are some special dislocation clusters, which are akin to the subgrain boundaries in crystals and the disclinations that border them. 
%They can be present either in dust clouds or in radiation beams.
  \end{enumerate}  

This paper, inspired for a large part by concepts elaborated in CMP, is intended to be read mostly by cosmologists. Within this scope, Sect.~\ref{cond-matt} is an overview of the theory of defects or singularities in CMP (the term 'defect' has a broader sense; there are non-singular defects); a short summary of the  literature on singularities in cosmology is presented Sect.~\ref{cosm-defects}. 

We recall in Sect.~\ref{cond-matt} that there are two categories of defects, (1)- those that come under the heading of algebraic topology (\textit{topologically stable defects}, which are but to a few exceptions true \textit{singularities} of the 'order parameter'), (2)- those
whose characteristics $-$ Burgers vector, rotation or Frank vector, either continuous or quantized $-$, are related to some element of the symmetry group. 
These \textit{Volterra defects} are 2D defects in spacetime, line defects in a 3D submanifold of the spacetime. 

We comment on defect theories in cosmology in Sect.~\ref{cosm-defects}:   
whereas Kibble's cosmic strings are related to the topological properties of {gauge} transitions, and Ruggiero \& al.'s defects to the topological properties of the Poincar\'e group, Hawking \& al.'s singularities do not relate at first sight to spacetime \textit{topological} properties, but their theoretical status relies on a deep analysis of the \textit{geometry} of causal 4D manifolds.

Section~\ref{poinforms} extends to $M^4$ (equipped with the Poincar\'e group $P(4)$) the concepts of defect (Volterra and topological defects) developed for $E^3$ (equipped with  the Euclidean group $E(3)$). In this latter case, defects are defined for crystals on the basis of their symmetry groups, which are subgroups of $E(3)$. 
The choice of the relevant subgroups of $P(4)$ to developing the same approach relies on the idea that the only subgroups of interest are those relating to massive particles or to radiation, in the sense given by \textcite{wigner39}. 
It is also justified by the nonexistence of a generalized cosmological principle befitting the 4D spacetime. 
The defects so defined are qualified of \textit{perfect defects}; they carry characteristics which relate to the elements of the corresponding subgroups. 

In Sect.~\ref{stab-cryst} our analysis runs into the definition of two sets of Minkowski's 'crystals', 
which are in fact geodesic congruences, timelike in the case of massive particles, null in the case of radiation. 
The \textit{physical reality} of these 'crystals' is discussed in Sect.~\ref{stronghyp}; one advanced argument is that it is correlated to the Hubble's expansion. 
Their characteristic defects (forms of matter and forms of radiation, as already mentioned) are then defined. 

The long Sect.~\ref{forms} is about Volterra forms, which all are continuous. \textit{Perfect} line defects in 3D spaces are presented in Sect.~\ref{forms-matt} and Sect.~\ref{forms-light}. 
Their role may be important in dynamical phenomena, precisely because of their continuous properties, by which it is meant that the line characteristics $\mathbf{b},\,\bm \Omega$ can be continuously varied. 
We also emphasize the existence of three classes of \textit{imperfect disclinations} that border misorientation walls for certain 4-momenta congruences in three dimensions, and are akin to the disclinations that border subgrain boundaries in crystals: \textit{imperfect $m$-disclinations}, \textit{imperfect $r$-disclinations}, and \textit{hyperbolic} (or \textit{boost}) \textit{disclinations}; 
they are of a paramount importance in the picture we propose of singularities in general relativity. 
Imperfect disclinations do not relate directly to the symmetry properties of the 'crystals', hence to the massive or massless nature of particles. 
We also speculate on the role played by continuous defects in the large scale structure of the Universe. 

A large part of Sect.~\ref{forms} might look somewhat technical to a reader little acquainted with the physics of defects in CMP; although the results which this section brings about are essential for a full understanding of continuous forms in $M^4$; it can be skipped in a first reading, except the short introduction, before \ref{forms-matt}, and the last part \ref{network}.

A fuller discussion of the results of the paper is presented Sect.~\ref{disc}. The investigation on the nature of the topological quantized forms, which all of them are $r$-defects, will be published later on. 

\section{Defects in Condensed Matter Physics}\label{cond-matt}
The concept of defect is rather well demarcated in CMP. 
There are two different approaches, which distinguish {\textsf{Volterra defects} (either continuous or quantized) and \textsf{topological defects}. 
 
\subsection{Volterra continuous and quantized defects}The underlying concepts were first developed by \textcite{volterra07}, in view to classify the singular solutions of Hooke's elasticity in an isotropic homogeneous solid (i.e. an amorphous material, also called a glass).  
Closed\textit{ line} singularities are from that point of view of a particular interest. 
The {Volterra process} (VP) which constructs them goes as follows \cite{friedeldisloc,klfr08}:  consider a surface $\Sigma$, drawn in a unstrained \textit{amorphous }specimen, bordered by a closed or an infinite line L$=\partial \Sigma$; the sample is then cut along $\Sigma$, and the two lips $\Sigma^+$ and $\Sigma^-$ of this \textsf{cut surface} $\Sigma$ are displaced relatively one to the other by a \textit{rigid} displacement ${\textbf{d}(\textbf{x})}=\textbf{b}+\bm \Omega(\textbf{x})$; $\textbf{b}$ is a translation and $\bm \Omega$ a rotation, $\textbf{x}$ is any point on $\Sigma$; L is an oriented loop, and the relative displacement of the two lips is determined by this orientation.
After addition of missing material in the void (if a void is created) or removal of superfluous matter (if there is multiple covering), the atomic bonds are reset on the lips and the specimen let to elastically relax. 
It can be shown that at the end of this process the singularities of the strain field are restricted to the line itself; the position of the lips is unmarked. 

A \textsf{dislocation} is a line defect that breaks a translational symmetry $\textbf{b}$, the \textsf{Burgers vector} of the dislocation; a \textsf{disclination} is a line defect that breaks a rotational symmetry generally represented by a vector $\bm \Omega$, called the \textsf{rotation vector} (or the \textsf{Frank vector}) of the disclination. 
The group of symmetry $E(3)\sim SO(3)\Box \mspace{1mu}\mathbb{R}^3$ of an euclidean manifold $E^3$ being continuous, the Volterra defects are \textit{continuous}, in a sense that will be made more precise Sect.~\ref{boundcond}.

Fig.~\ref{fig01.pdf} illustrates various possible dislocation or disclination lines in function of the orientation of the Burgers or rotation vectors with respect to the line L, assumed here to be an infinite straight line.
\begin{figure}
%\begin{center}
\includegraphics[width=3. in]{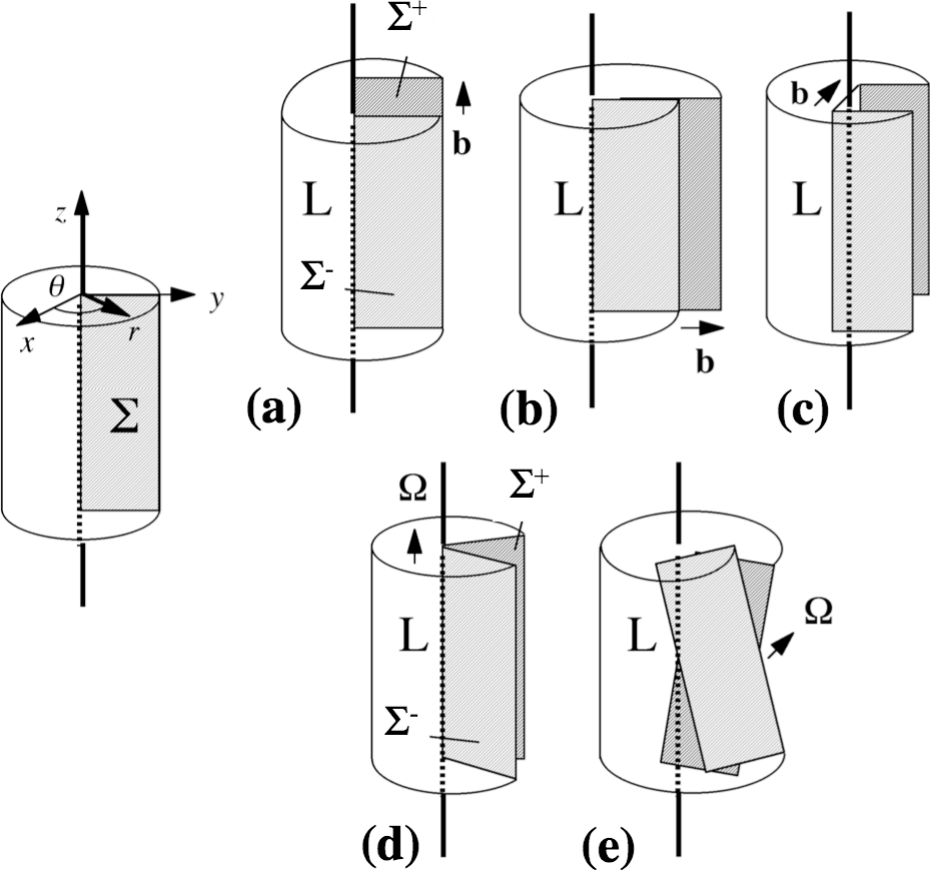}
   \caption{ Schematic representation of the Volterra process for a straight line that borders a half plane cut surface $\Sigma$. (a) screw dislocation; (b, c) edge dislocations: these two processes that relate to the same dislocation (but to a global $\mathrm{\pi}/2$ rotation) yield the same dislocation in an isotropic medium, after relaxation of the stresses; (d) wedge disclination; (e) twist disclination. \textit{adapted from} \cite{klfr08}.}
 \label{fig01.pdf} 
%\end{center}
\end{figure}

This Volterra process can be extended with the same result to an \textit{ordered} medium, provided $\textbf{d}$ is an element of symmetry of the system. 
In that case the line defects are \textit{quantized} if $ H$ (the {Bravais group}) is discrete, as in a crystal.
Since $\textbf{d}$ is an element of symmetry $\in H$, there is no singularity of the order parameter along the lips, after completion of the process; see ref. \cite{friedeldisloc} for details. 

VP applies to ordered media like crystals \cite{friedeldisloc}, liquid crystals \cite{kln}, or quasicrystals \cite{kle88}, i.e. to cases where the elements of the symmetry group $H $ correspond to possible operations of symmetry in space ($H$ is a subgroup of the space group $G=E(3)$)\footnote{the case of quasicrystals is rather special; the symmetry group $G$ is a $E(n)$ with $n>3$.}; 
however line defects can also be considered in other types of ordered media, where the Volterra process does not makes sense, e.g. vortex lines in superconductors and superfluids: the group of symmetry is then a {gauge group} $H$, which itself breaks some gauge symmetry group $G \nsim  E(3)$. $H$ as a gauge group is not a subgroup of $E(3)$, as a rule. 
But there is a \textsf{unique} theoretical frame in which quantized Volterra and gauge line singularities both appear as defects, as follows. 

\subsection{Topological defects}\label{topdefa} They are classified by the non-trivial elements of the homotopy groups $\pi_{i}(\cal{M})$ of the manifold $\mathcal{M}=G/H$, which is the set of left cosets $gH\,(g \in G)$ of the symmetry group $H \subset G$. $H$ is also called the \textsf{little group} or the \textsf{isotropy group}; 
$\cal{M}$ is the \textsf{coset space}, also called the manifold of degenerate vacuum or the order parameter space \cite{toulouse76}. 
The topological classification of defects extends to any physical medium with a symmetry group $G$ broken to $H$. $H$ is either the {symmetry group} of the crystal, or a group encompassing more subtle symmetries, like magnetic or electric properties of various kinds, or a {gauge group}, as stated above. 

A considerable advantage of the topological approach is that it extends to defects of any dimensionality (wall defects are classified by $\pi _{0} (\cal{M})$, line defects by $\pi _{1} (\cal{M})$, point defects by $\pi _{2} (\cal{M})$, configurations\footnote{also called \textit{textures} but this is a misnomer; the term \textit{texture} has since long a well defined meaning in CMP: it designates an element in the category of the figures made of a irrreducible (in some sense) set of defects.} by $\pi_{3}(\cal{M})$). For reviews see \cite{michel,mermin,trebin}. 
However, whereas the topological classification provides criteria relevant to the line defect \textit{topological stability}, their \textit{energetic stability} is better discussed in the frame of the Volterra process. %, only their \textit{topological stability}, (i.e. it classifies defects that cannot disappear without breaking the topological texture of the specimen).  
Furthermore the Volterra approach investigates not only quantized line defects belonging to the non-trivial classes of the fundamental group $\pi_{1}(\cal{M})$ (i.e., topologically stable), but also the continuous defects alluded to previously and that can be topologically assigned to the trivial class $\{1\}\in \pi_{1}(\cal{M})$ (not topologically stable). 
These latter singularities play an important role in the mobility and flexibility of quantized defects \cite{klfr08}.\\

 In order to recognize the homotopy class $\in \pi_{i}(\cal{M})$ of an actual  topological defect $\Lambda _i$, one surrounds it by a submanifold $\Gamma_i, \,\dim \Gamma_i = i, \,\Lambda _i \cap \Gamma_i =\emptyset$, e.g. a closed loop $\Gamma_1$  surrounding a line $\Lambda _1$ for $i = 1$ (the \textsf{Burgers circuit}), a surface with the topology of a 2-sphere for $i = 2$, etc. 
 If the defect belongs to a space of dimensionality $n$, then $\dim {\Lambda _i} = n-1-i$. 
 The little group at each point of $\Gamma_i$ is mapped onto $\cal M$; the homotopy class of this image is the class one is looking for \cite{toulouse76}.  
The $\Gamma_i$s are generalized Burgers circuits; we call them \textsf{Burgers surrounds}. 

\subsection{Habit manifolds and dynamical manifolds}\label{manifolds} In both approaches one is led in CMP to consider an Euclidean three-dimensional manifold \textit{filled} by a \textsl{substance} (matter, in practice), whose euclidean character is preserved whatever the defect content may be. 
In that sense it is a \textsf{habit manifold}, which provides a reference frame. Another manifold has been introduced in CMP, endowed with local \textit{torsion} carried by dislocations, and local \textit{curvature} carried by disclinations present in the sample; for a review see \cite{kroner}. 
The mapping of an ordered medium and its defects onto such a \textsf{dynamical manifold} with variable torsion and curvature works insofar as the defects can be constructed by a Volterra process. 
The topology of such a manifold depends on its defect content. The analogy with the \textsf{spacetime} of General Relativity, whose metric $ds^2=g_{\mu\nu}dx^\mu dx^\nu$ represents a variable distribution of curvature and torsion, has often been noticed \cite{hehl}.

 \section{Defects in cosmology; a summary of the present situation} \label{cosm-defects}

\subsection{{Hawking$-$Penrose} singularities} 
{Hawking} and {Penrose} have introduced spacetime singularities of null and time-like geodesics. 
They appear as endpoints of \textsf{incomplete geodesics} and are present in the core structure of Black Holes and in the 'initial singularity', the Big Bang.

The question arises whether these singularities can be defined as Volterra or topological defects in the sense discussed above. The situation would therefore be somewhat analogous to that met in a distorted director field, for example in a nematic.  
It is known that the envelopes of the integral curves of the director field\footnote{A \textsf{director}, in the terminology of the physics of soft condensed matter, is a vector $\textbf{n}$ physically equivalent to its opposite $\textbf{n}\sim -\textbf{n}$. 
The long axis of the calamitic molecules which are the building blocks of the nematic phases can be represented by a director \cite{pgg74}.} are special realizations of topological or Volterra defects, see Sect.~\ref{congeucli}. See also Sect.~\ref{boundcond} about the analogs of incomplete geodesics in CMP. Another analogy is with focal conic domains (FCDs) in smectic phases, which are well-known caustic singularities in liquid crystals \cite{kln}.

If it is so, HP singularities should be akin to disclinations of \textit{physically recognizable} null or time-like geodesic families. This question is not set.

  \subsection{Kibble cosmic strings} 
  The phase transitions between different types of particles, except if they occur on coherence lengths of the order of the universe size, are necessarily attended by the production of defects of the type foreseen by \textcite{kibble76}. The notion of \textit{topological} defects in cosmology is contemporaneous with that one in CMP \cite{toulouse76,rogula76}; they were published independently and remained rather long mutually ignored.
 
  Cosmic strings are formed in many symmetry-breaking phase
transitions from a group $G$ to a subgroup $H$. As usual, the homotopy groups of the manifold of degenerate vacuum $\mathcal {M} =G/H$ determine the types of defect that can form.  
For instance, Grand Unified Theories (GUTs) have been proposed based on simple groups $G$ such as $SU(5)$ or $SO(10)$, or on semisimple groups such as the left--right symmetric $G=  SU(4) \times SU(4)$, broken in one or more stages from $G$ down to $G_{321}= SU(3) \times  SU(2)  \times U(1)$. 
The spacetime of general relativity is, according to the standard cosmological model, the place where such phase transitions between hot {Planck} particles and cold contemporaneous elementary particles (photons, leptons, quarks in confined assemblies, bosons, etc) do occur. The symmetries at work are {gauge symmetries}.
 
  The presence of {cosmic} strings has never been documented, but hope still prevails \cite{kibble-2004}.
   
  \normalsize
   
\subsection{Poincar\'e -Volterra defects} \label{PVVP}
We shall call \textsf{Poincar\'e$-$Volterra defects} the Volterra line defects of a GR spacetime $-$ here we restrict to $M^4$ $-$ considered as the habit manifold of some amorphous substance. 
$M^4$, which has everywhere zero curvature and torsion, is globally invariant under the continuous Poincar\'e group $P(4)=L_0 \Box \mathbb{R}^4$, and acquires curvature and torsion in the presence of line defects the same way a 3D Euclidean space does when it is the habit space of a plastically deformed material \cite{kroner}, see Sect.~\ref{manifolds}. $L_0$ is the connected Lorentz group.

Thus Poincar\'e$-$Volterra defects are constructed by employing the analogue of VP for $M^4$. 
The displacement of a point $\textbf{x} \in  M^4$ belonging to the cut manifold of some defect now reads $d^a(\textbf{x})={b^a}+{L}(\textbf{x})\negthinspace:\negthinspace \textbf{x}$, where ${b^a}$ is still a 4D translation and ${{L}}(\textbf{x})$ a Lorentz transformation. 
The continuous \textit{disclinations}, which locally break $L_0$, carry {curvature} and can therefore be interpreted in the {Einstein} spirit as dense clusters of massive particles. The continuous \textit{dislocations}, which locally break $\mathbb{R}^{4}$ and carry {torsion}, can be interpreted in terms of spin-momentum density, but this interpretation is far from being clear. 
After the introduction of the defects, the final resulting spacetime carries torsion and curvature \textit{densities} defined by the metric tensor $g_{\mu\nu}$. These $g_{\mu\nu}$ are equivalent to 3D strains.
 
 Defects of the kind discussed here have been investigated, to quote a few: spinning strings with cosmic dislocations, pure spacetime dislocations, chiral strings, etc. A large part of the literature on the subject is cited and commented in \cite{ruggiero03,puntigam97}.
 
\section{Poincar\'e forms}  \label{poinforms}
 The Poincar\'e$-$Volterra continuous defects of Sect.~\ref{PVVP} are those of a glass in $M^4$. We shall argue that the existence of the full group of defects classified by the elements of $P(4)$ contradicts the cosmological principle. 
But some of its defect subgroups do not incur this criticism; the remainder of this paper is devoted to their analysis. 

A flat $M^4$ is the habit manifold of photons that can be split into null geodesic sets of \textit{parallel} lines, or of massive particles also split into time-like geodesic sets of \textit{parallel} lines. The defects of these \textsf{ideal congruences} affect also the metric, and thus $M^4$, according to the Einstein's equation. 
We call these defects \textsf{Poincar\'e forms}. Any distorted, defective, congruence results from an ideal  congruence in which Poincar\'e forms are introduced. 
 
 \subsection{The Poincar\'e group \textit{P}(4): a reminder}\label{PGr}
  
 We are interested in the connected component $L_0$ of the Lorentz group, whose elements preserve both orientation and the direction of time. 
 $\mathbb{R}^{4}$ is the abelian group of translations in $M^4$, it is a normal subgroup of the Poincar\'e group and thereby of its connected version $P(4)=L_0 \Box \mspace{1mu}\mathbb{R}^{4}$. 
 Because $L_0$ is not simply connected, (its fundamental group is $\pi_1(L_0)\sim \mathbb{Z}_2$), we introduce its double cover $\widetilde{L_0}$ which as a manifold is simply connected, $\pi_1(\widetilde{L_0})\sim \{1\}$, and whose use proves simpler. $\widetilde{L_0}$ is isomorphic to the special linear group $S\ell (2,\mathbb{C})$, which is the group of 2$\times$2 matrices 
$\begin{array}{|cc|}
  a    &   b \\
   c   &   d \\
\end{array}$
with complex entries and unit determinant $ad-bc=1$. 

With these tools at hand, a Lorentz transformation is then expressed as follows \cite{wigner39}. Let $x^a=\{x^0,x^1,x^2,x^3\}$ {be an event} $\in M^4$, we represent it by the 2$\times$2 hermitian matrix
\begin{equation}%{multline}
\label{e1}
{{x}}=x^0 e +x^1 \sigma _1+x^2 \sigma_2+x^3 \sigma _3 %\\ 
=\begin{array}{|cc|}
 x^0+x^3     &  x^1-ix^2  \\
  x^1+ix^2    &   x^0-x^3
\end{array}\,
\end{equation}%{multline} 
($ e$ is the unit matrix, the $ \sigma_i$'s are the Pauli matrices); 
the Lorentz transformation of $x$ under the action of a matrix $A \in Sl (2,\mathbb{C})$ can then be written:\begin{equation}
\label{e2}
x\mapsto x'=AxA^*,
\end{equation}where $A^*=\overline{A^t}$ is the conjugate transpose of $A$. 
Notice that $A$ and $-A$ yield the same result, which heals the possible problems raised by the 2:1 homomorphism (the \textsf{spinor map}) introduced above \begin{equation}
\label{e3}
\mathrm{Spin}:\,S\ell(2,\mathbb{C})\rightarrow {L_0}.
\end{equation} 
Notice also that in Eq. \ref{e2}, \begin{equation}%{multline}
\label{e4}
\mathrm{det}(x)\equiv (x^0) ^2 -(x^1) ^2-(x^2) ^2-(x^3) ^2=-\|{x^a}\|^2%\\ 
=-\|{x}^{'a}\|^2=\mathrm{det}(x'),\end{equation}%{multline} 
(see the end of this section for the notations we use.) 
Let us recall that the norm $\|{x}^a\|^2=x^\alpha\,x_\alpha $ is $<0$ for ${x}^a$ timelike, $>0$ for ${x}^a$ spacelike, $=0$ for ${x}^a$ null.
Elements of $Sl (2,\mathbb{C})$ are also called \textsf{spin transformations}: a $2\pi$ rotation in $L_0$ cannot be continuously deformed to the identity transformation, whereas a $4\pi$ rotation can. 

Eq.~\ref{e3} maps the unitary subgroup $SU(2)\subset S\ell(2,\mathbb{C})$ onto the rotation subgroup $SO(3) \subset L_0$. 
 An element $A\in SU(2) \sim \widetilde{SO(3)}$ can be written in matrix form $ \begin{array}{|cc|} a     &  b  \\  -\bar{b}    &   \bar{a} \end{array},\, \|a\|^2+\|b\|^2=1$.\\

 \small{\textit{Our notations.} {We shall use the term \textsl{space} to denote a spacelike submanifold of $M^4$, e.g., a 3-space. 
 Usually, a bolded symbol will refer to a 3-space vector $\textbf d = \{ d^1,d^2,d^3\}$; $d^a$ refers to a vector in $M^4$ $-$ with components $d^\alpha,\, \alpha= 0,1,2,3$, e.g., $d^a=\{ d^0,d^1,d^2,d^3\}$ $-$ and $d_a=\{ d_0,d_1,d_2,d_3\}$ to the dual vector. 
 A point in a 3-space or in $M^4$ is bolded, e.g., $\textbf x$. We also use the notation ${{d}}$ for $d = d^0 e +d^1 \sigma _1+d^2 \sigma_2+d^3 \sigma _3$, as in Eq.~\ref{e1}.} \normalsize
 
 \subsection{Singularities of a glass in  $E^3$ and in $M^4$: the generalized cosmological principle}\label{sing-glass}
Let us come back to a distinction we have already introduced. 
There is no concept of \textit{breaking} of symmetry that is \textit{natural} to an \textit{empty} manifold; the situation is different if this manifold is the actual {habit manifold} of some kind of \textsf{substance}, i.e. for a physicist: \textsf{energy}, materialized or not. 
We introduce the term of 
\textit{substance} to connote phases of massive particles, e.g., baryons, {as well as} of massless ones, e.g., photons. 
 
$-$ To make things intuitively simple, consider first a 3D Euclidean space $E^3$. 
  Is it physical to interpret its symmetry group ${E(3)}$ as the group of symmetry $H$ of some substance residing in $E^3$?  The coset space of the considered substance would then be {$E(3)/H$}, i.e., in the present case $\mathcal{M}_{3}=E(3)/E(3)=\{1\}$. 
  There are no topologically stable defects; the only possible defects are continuous line singularities, which can be constructed by a Volterra process. Such a substance, called a \textit{amorphous medium} or a \textit{glass}, has a physical reality in CMP.

$-$ Consider now the $M^4$ spacetime as a habit manifold endowed with the symmetry group $P(4)=L_0 \Box \mathbb{R}^4$.
If $M^4$ is filled with a substance that has the same properties of invariance as itself, a \textit{spacetime glass} so to speak, the situation is comparable to the previous one: $\mathcal{M}^4=\{1\}$ and the admissible defects are continuous and constructed by Volterra processes involving any of the $P(4)$ symmetries, cf. Sect.~\ref{PVVP}.  
Since all the events in a spacetime glass are equivalent under the action of ${P(4)}$, such a situation implies the \textit{generalized cosmological principle} employed by \textcite{bondi48} and \textcite{hoyle48} in their models of a steady state universe. 
But there is little doubt to-day that this principle is not obeyed, and that one has to be content with a \textit{restricted cosmological principle}; at cosmic time $\tau$, the space alone is homogeneous and isotropic, at least on galactic supercluster scales. 

Hence: \textit{there is nothing like a \emph{spacetime glass}, thereby nothing like the Poincar\'e$-$Volterra defects it would carry}. It is improper to consider ${P(4)}$ as the group of symmetry of a substance residing in $\mathcal{M}^4$.

\subsection{Minkowski's crystals: the restricted cosmological principle}\label{Mcrystrest}
  The abandonment of the generalized cosmological principle forces us to consider substances in $M^4$ arranged with symmetries belonging to the subgroups $H$ of $L_0$. 
  We have the relation $\mathcal{M}=L_0/H=\widetilde{L_0}/\widetilde{H}=S\ell(2,\mathbb{C})/\widetilde{H}$, so that the topological classification of defects by the homotopy groups $\pi_i(\mathcal{M})$ requires only the knowledge of the double covers $\widetilde{H}$.  
  
  \textcite{wigner39}, in his epoch-making article \textit{On the Unitary Representations of the Inhomogeneous Lorentz Group} (the connected Poincar\'e group), has demonstrated the equivalence between the unitary irreducible representations of some subgroups of $P(4)$ and the elementary particles.   
  Following him, we define at the present level of analysis an elementary particle in $M^4$ by a 4-momentum ${p}^a=\{p^{0},p^{1},p^{2},p^{3}\}$ alone; we further assume that a spacetime in an ideal state is filled homogeneously with the same particles carrying the same momentum. 
  This is featuring a \textit{congruence of geodesics}, invariant under some subgroup of $P(4)$, thought of thereby as a \textsl{crystal} in $M^4$. It is sufficient to record one 4-momentum ${p}^a=\{p^0,p^1,p^2,p^3\}$ at some point $\textbf{x} \in M^4$, and the density at rest $\rho$, to know the entire $M^4$-crystal, Fig.~\ref{fig02.pdf}. 
  We come back later on to the physical $vs$ fictitious character of $M^4$-crystals, Sect. \ref{stab-cryst}. But observe that this analogy suggests an approach to singularities in spacetime physics in congruity with the approach to crystal singularities in CMP, i.e., somewhat different from that one of Hawking \& Penrose but hopefully, yielding results consistent with theirs. 
  
 There are four types of orbits described by the 4-momentum vectors $p^a$ under the action of $L_0$, depending on the sign and value of the norm $\|{p^a}\|^2$. 
 Each type is characteristic of a different type of particles, in our terminology different substances, that fill homogeneously the entire spacetime. These are:
 \begin{figure}
%\begin{center}
\includegraphics[width=2. in]{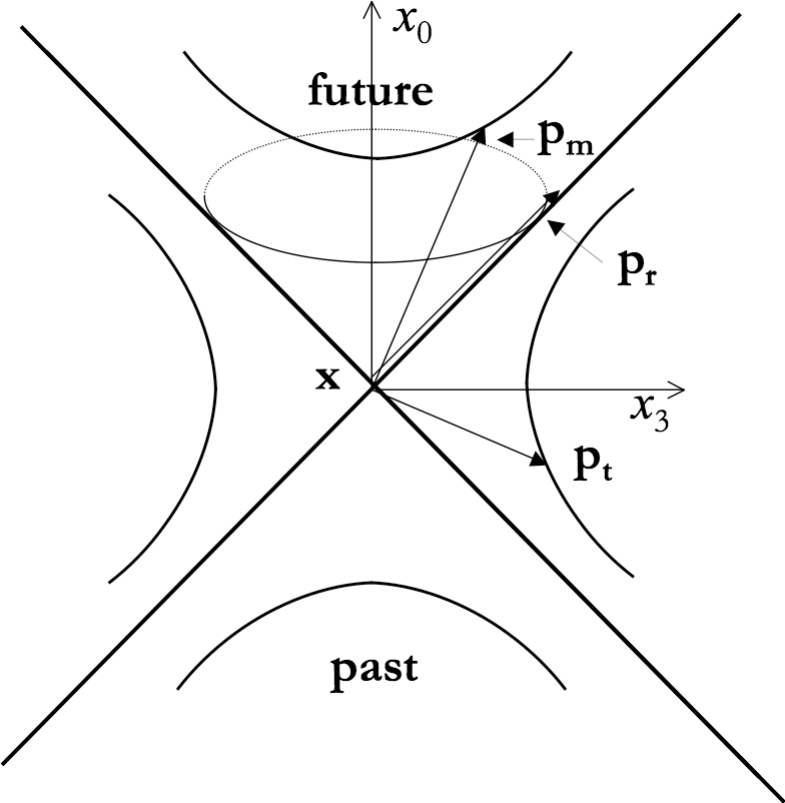}
   \caption{ 3D representation of the light cone at some event $\textbf{x}$, cut in the $\mathbf{e}_0\,\mathbf{e}_3$ plane. 
   The three Wigner's categories of particles are represented by the vectors ${p}^a_m$ (timelike), ${p}^a_r$ (null), ${p}^a_t$ (spacelike) and the corresponding orbits, which are hyperboloids, or the light cone with the origin excluded. 
   A cross section of the future light cone is drawn; it is a 2-sphere in 4D. The origin ${p}^a=0$ is a 4$^{rth}$ category in itself, but of little interest.}
 \label{fig02.pdf} 
%\end{center}
\end{figure}

   \begin{enumerate}
   \item massive particles, parameterized by their mass $m^2>0,\, m^2=-\|p^a_m\|^2$ and their spin $s=0,\,\frac{1}{2},\,1,\,\frac{3}{2}, \,\mathrm{etc}$ (but we consider as the same particle those with different spins).
    The little group $H_m$ that leaves ${p}_m^a$ invariant is isomorphic to the group of 3D rotations; $H_m\sim {SO(3)}$, with $\widetilde{H_m}\sim SU(2)$. 
    The orbit belongs to a hyperboloid with two sheets (one in the past, the other in the future). In Fig.~\ref{fig02.pdf}, where we have taken $p_{0}>0$, it is the future sheet,

    \item massless particles, parameterized by their mass $m=0$, hence $\|p^a_r\|^2=0$ and their helicity $s=0,\,\pm1,\,\pm2, \,\mathrm{etc}, $ (but we consider as the same particle those with different helicities).
     The little group $H_r$ that leaves ${p}_r^a$ invariant  is isomorphic to the 2D Euclidean group; $H_r\sim {E}(2)= SO(2)\Box \mspace{1mu}\mathbb{R}^2 $, the semidirect product of a circle by a plane; its double cover in $S\ell(2,\mathbb{C})$ is denoted $\widetilde{H_r}\sim \widetilde{E(2)}$, but is isomorphic to ${E}(2)$.
     \footnote{The $SO(2)$ component of $E(2)$ can be represented by a circle, with angles $0\leqslant \theta<2\pi$. This circle is covered twice in $\widetilde{E(2)}$, with $0\leqslant \theta<4\pi$.} 
     The orbit is one of the sheets of the light cone (the origin $x=0$ is excluded), in Fig.~\ref{fig02.pdf} the future half-cone,
  \item massive particles, parameterized by their mass $m^2<0,\, m^2=-\|p^a_t\|^2$ and their spin $s=0,\,\frac{1}{2},\,1,\,\frac{3}{2}, \,\mathrm{etc}$; the little group that conserves ${p}^a_t$ is isomorphic to ${Sl(2,\mathbb{R})}$; the orbit is the spacelike hyperboloid with one sheet,  
  \item null particles $p^a=0$: the little group is the full proper Lorentz group $L_0$.
\end{enumerate}

   The last case is uninteresting, it would bring us back to a spacetime devoid of substance, which we have eliminated. The 3$^{rd}$ case describes \textsf{tachyons}, i.e. particles with a velocity larger than the velocity of light. Therefore only the first two cases are interesting. They represent respectively:
\begin{enumerate}
  \item a $M^4$-crystal made of an homogeneous dust of matter ($m$-crystal) with parallel 4-momenta, $H_m\sim {SO(3)}$,
  \item a $M^4$-crystal made of an homogeneous radiation ($r$-crystal), with parallel 4-momenta, $H_r\sim {E}(2)$.
 \end{enumerate}   
 
Again, the notion of singularity applies only if it concerns the singularity of a well defined substance. 
In the case of a homogeneous massive substance, the field of 4-momenta is tangent to a \textit{congruence of parallel timelike geodesics}. In the case of a homogeneous radiative substance, the field of 4-momenta is tangent to a \textit{congruence of parallel null geodesics}. 
These are our $M^4$-crystals in their ideal state. According to Wigner's classification, there are only two types of $M^4$-crystals, each type forming a continuous set whose elements differ by their masses, energies, spin, helicity.

  \section{The physical nature of $M^4$-crystals: their defects} \label{stab-cryst}
 \subsection{Congruences in an Euclidean space $E^3$}\label{congeucli}
The classification of the singularities of a congruence of lines in $E^3$ is usually approached by the methods of differential geometry. 
Another approach is topological; it considers the singularities of a congruence as the defects of a set of parallel lines, considered as forming a crystal in $E^3$, i.e., as the family of possible singular distortions of such a congruence that do not break the congruent character of the lines, except at the singularities themselves \cite{kleman73b}. 
This is illustrated by an infinite \textit{liquid} ferromagnet, whose  ground state can be represented by an \textsl{oriented} congruence of parallel magnetization integral lines. Such a congruence has topological properties described by the coset space $S^2\Box \mspace{1mu}\mathbb{R}^{3}$; hence line defects are not topologically stable $-$$\pi_{1}(S^2)\sim \{1\}$ $-$ but may exist as Volterra defects, whereas there are stable point defects $-$$\pi_{2}(S^2)\sim \mathbb{Z}$. 
The coset space of a nematic phase, which is a congruence of \textsl{unoriented} lines, is $P^2\Box \mspace{1mu}\mathbb{R}^{3}$, where $P^2= S^2/\mathbb{Z}^2$, the projective plane. 
We still have the same point defect classification $\pi_{2}(P^2)\sim \mathbb{Z}$, but the line defect classification is richer, since apart the still conceivable Volterra defects (attached to $\mathbb{R}^{3}$ and to continuous rotations about the line axes), one has $\pi_{1}(P^2)\sim \mathbb{Z}^2$.

In these two examples the congruences in their ideal state are made of geodesics, but lose that latter characteristic when they are deformed and acquire singularities. 
This modification is governed by the laws of elasticity. In GR the singularities, governed by Einstein's equations, get endowed with curvature and torsion, the same concepts used to describe singularities in CMP, Sect.~\ref{manifolds}. 
Our definition of singularities in $M^4$, akin to that one for $E^4$ crystals, ensures that the curvature and torsion have a similar origin in both cases.

 \subsection{Topological and Volterra forms in $M^4$}\label{topcong} 
   We have in $M^4$ two types of relevant geodesic congruences. Their singularities are of a different nature. We call them \textsf{forms of matter} (in a $m$-crystal) and \textsf{forms of radiation} (in a $r$-crystal).

 \subsubsection{Timelike geodesic congruences}\label{tgc} 
 Their coset space $\mathcal{M}_m$ is topologically equivalent to the orbit of a massive particle, $\|{p}_m^a\|^2=-m^2$, represented Fig.~\ref{fig02.pdf} as the future sheet of an hyperboloid with two sheets, i.e., $\mathcal{M}_m=H^3$. 
 This coset space is also obtained as the quotient $L_0/SO(3)=H^3$, whose little group $H_{m}\sim { SO(3)}$. The corresponding homotopy groups are trivial
\begin{equation}
\label{e5}
  \pi_{i}({H^3)\sim \{1\}}.
\end{equation} 
There are no topologically stable defects for timelike congruences, i.e., for massive particles.

We are left with Volterra defects. $H_{m}$ is the point group; the full group of symmetry, translations and rotations included, is $ P_{m} \sim H_m\Box \mspace{1mu}\mathbb{R}^{4} \sim SO(3)\Box \mspace{1mu}\mathbb{R}^{4}$. 
We call  the corresponding forms: matter-dislocations $-$in brief \textsf{$m$-dislocations}$-$ and \textsf{$m$}-\textsf{disclinations}. They are classified respectively by the elements of $\mathbb{R}^{4}$ and of $SO(3)$. 

 \subsubsection{Null geodesic congruences} \label{ngc} Their coset space $\mathcal{M}_r$ is  topologically equivalent to the future lightcone with apex removed, $ \|{p}_r^a\|^2=0$, i.e., $\mathcal{M}_r=S^2 \times \mathbb{R}$. 
 This is the cartesian product of a 1D generator of the future null cone (with the apex removed) and a non self-intersecting cross section of this cone, (one is drawn in Fig.~\ref{fig02.pdf}), which has the topology of a 2-sphere. 
 The little group $H_r$ being $E(2)$, this coset space can also be obtained as the quotient $L_0/E(2)$.
The corresponding homotopy groups are \begin%{multline}
{equation}
\label{e6}
  \pi_{i}(\mathcal{M}_{r})\sim \pi_{i}(S_{2}) \rightarrow \\ 
   \pi_{1}(\mathcal{M}_{r})\sim \{1\},\,\pi_{2}(\mathcal{M}_{r})\sim \mathbb{Z},\,\pi_{3}(\mathcal{M}_{r})\sim \mathbb{Z}.
\end%{multline}
{equation}
$\mathbb{Z}$ is the group of integers. Hence there are topologically stable defects for radiation congruences, whose dimensionalities in $M^4$ are: 1 for $\pi_{2}(\mathcal{M}_{r})$ and 0 for $\pi_{3}(\mathcal{M}_{r})$.

Line defects are Volterra defects, since  $\pi_{1}(\mathcal{M}_{r})\sim \{1\}$.  
These line defects are continuous, since the group of symmetry of a massless substance, $P_{r}\sim E(2)\Box \mathbb{R}^{4}$, is continuous. 
We call the corresponding forms: radiation-dislocations $-$in brief \textsf{$r$-dislocations}$-$ and \textsf{$r$}-\textsf{disclinations}. 

 \subsubsection{Reality of the singularities of congruences, philosophical argument and empirical evidence}\label{stronghyp}
The little group of a liquid ferromagnet, $SO(2)$, differs from the little group ${SO(3)}$ of massive particles, although both congruences have similar topologies.  Hence the singularities of a timelike congruence in $M_4$, the $m$-forms, differ from the Volterra singularities of an ordinary congruence in 3-space. Such a non-equivalence appears  even more clearly if one considers the elusive case of tachyons; their related congruence is spacelike, and so is the congruence of a liquid ferromagnet. However the little groups cannot be more different, $SO(2)$ for a liquid ferromagnet, ${Sl(2,\mathbb{R})}$ for tachyons. This remark emphasizes the physical importance of the substance that lives on the congruence, despite the fact that the two related congruences have the same representation in a 3-space. And as far as $r$-forms are concerned, they differ from Volterra defects in a ferromagnet although the coset spaces are the same $-$whereas the related congruences are somehow comparable.

The Volterra forms of $m$-congruences and $r$-congruences will be discussed in more details in Sect.~\ref{forms-matt} and \ref{forms-light}. %, and the topological forms of $r$-congruences in Sect.~\ref{toprad}. 
 The interest of such investigations rely on a strong hypothesis, namely that these $m$- and $r$-defects correspond in a \textsl{physical} sense to the singularities of the corresponding $m$- and $r$- substances. One is indeed led to believe \textsf{by induction} from the foregoing remarks that it is so, as it is difficult to imagine that such differences, \textit{since they exist}, would not have a physical significance (an induction which, as the alert reader may have observed, has some kinship with Anselm's ontological argument). 
 
 But one may prefer a seemingly stronger argument that qualifies these congruences as veridical $m$- and $r$- crystals; we believe that the \textit{empirical evidence of the expansion of the Universe} provides such an argument. Indeed, as noticed incidentally by \textcite{dirac38}, the reality of Hubble's expansion means that \textsl{"all the matter in a particular part of space has the same velocity (to a certain degree of accuracy) and suggest\emph{[s]} a model of the universe in which there is a \textsl{natural velocity} for the matter at any point$\cdots$ \emph{[which]} provides us with a \textsl{preferred time-axis} at each point, namely, the time-axis with respect to which the matter in the neighborhood of the point is at rest"}. Following these remarks, let $v_r$ be the Hubble velocity at some point at a distance $r$ from some observer. A particle of mass $m$ (which we specify here) has a 4-momentum $p_m=-m\{\gamma,\gamma v_r, 0,0\}$. One can imagine that all the particles of matter of a certain type, in a given cell of space and in free fall, have to a certain degree of accuracy the same 4-momentum, so that all types of particles have parallel 4-momenta. Thereby the 'orientation' of this cell, considered  as the habit cell of a $m$-crystal, possibly made of particles of different masses $-$ an 'alloy', so to speak $-$ is well defined. This picture of the Universe, made of different cells slightly misoriented one with respect to the other, is in accordance with our present understanding of its large-scale structure, as we shall argue in Sect.~\ref{network}, where we come back to this topic, and in the discussion, Sect.~\ref{disc}.
 
 Dirac's natural velocity is in agreement with the idea that the $m$-momentum (and the 3-space velocity) may fluctuate from a particular part of space to another, whereas radiation particles follow unperturbed the same null geodesic.%We forget this difficulty for the time being and develop the question of $m$-crystals .

 \subsection{Forms in $M^4$ $vs$ defects in $E^3$}\label{boundcond}
 It is generally agreed that a GR singularity is carried by an \textit{incomplete geodesic}, meaning that such a geodesic runs into a \textit{hole} of the spacetime in a finite amount of time (or a finite variation of the affine parameter for a null geodesic) $-$ in which hole the spacetime is not properly defined $-$,
 or took birth in a hole a finite time ago, see \cite{wald84}, chapter 9.  No doubt that \textsl{topological} $r$-forms are accompanied by such singularities; the question arises whether it is the same for \textsl{continuous} Volterra defects. 

The foregoing definition of a spacetime singularity is perfectly compatible with the image we have of a \textsl{topological} Volterra defect in CMP, i.e., when the defect belongs to a non-trivial class of $\pi_1(\cal M)$; the VP can be smoothly achieved on the cut surface $\Sigma$, but has to stop short at a small distance of the line L$=\partial \Sigma$ itself, where it is not defined. It thus appears an incomplete row of atoms $-$ the equivalent of an incomplete geodesic; the usual elasticity equations are not valid in a small \textsf{core} region about the line $-$ the equivalent of the hole. The core region has the topology of a toric cavity L$\times S^1$. Fig.~\ref{fig03.pdf} represents a straight \textsl{edge} dislocation which displays such incomplete rows of atoms. 
    \begin{figure}[h]
%\begin{center}
\includegraphics[width=1.5 in]{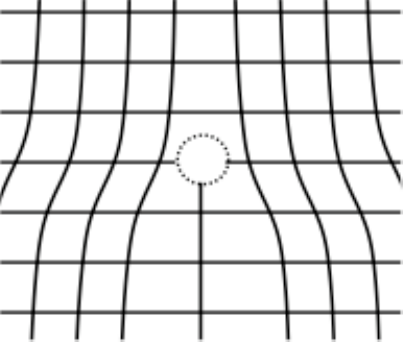}
   \caption{ 2D cut orthogonal to an edge dislocation line L in a square lattice. The line intersections represent atoms. Observe the presence of incomplete rows of atoms hitting the core region (the interior of the dotted circle) where the 'order parameter' of the crystal is no longer defined.}
 \label{fig03.pdf} 
%\end{center}
\end{figure}

Consider now a \textsl{continuous} Volterra defect which belongs to the trivial class of $\pi_1(\cal M)$; the Burgers circuit would not measure a nil displacement vector, but the core can vanish away if this displacement vector dissipates into Volterra defects with infinitesimally small displacements, which is topologically allowed and physically feasible by an irreversible viscous dissipation process in CMP. At low temperatures this process might be little activated, and a singularity still be visible. It is not known if a comparable spread of the core exists in GR. Actually, the gravitational forces may act in an opposite direction and (meta)stabilize a core singularity for a continuous defect, also accompanied with an entropy increase!

  \section{Continuous forms} \label{forms}
This section provides a few highlights on the characteristic 'distortions' brought by the various types of \textit{continuous forms} in $M^4$, which are all described succinctly.  These continuous forms are the forms classified in the Volterra manner by the elements of the little groups, and the boost-disclinations. We do not go as far as the calculation of the metric tensor $g{_{\alpha \beta}}$; this can be handled with the help of 
ref. \cite{puntigam97}, which gives solutions for a number of Poincar\'e forms that can be adapted to the examples we provide. We recall that there are two types of relevant crystals in $M^4$:

\textit{$m$-crystals. }$P_{m}\sim SO(3)\Box \mspace{1mu}\mathbb{R}^{4}$ is the group of symmetry of \textit{timelike congruences}. Therefore, in the corresponding Volterra process, $\mathbb{R}^{4}$ is the set of the Volterra dislocation invariants and $H_m \sim SO(3)$ the set of the Volterra disclination invariants.
     
     \textit{$r$-crystals. }The situation is somewhat more complex for \textit{null congruences}. $P_{r} \sim E(2)\Box \mspace{1mu}\mathbb{R}^{4}$ is the group of symmetry of \textit{null congruences}; $H_r \sim E(2) = SO(2)\Box \mspace{1mu}\mathbb{R}^2$. In terms of the Volterra process, $E(2)$ is the set of disclination invariants, which we loosely call \textsf{$r$-rotations}, or simply \textsf{rotations} when there is no risk of confusion, by analogy with the usual case; $\mathbb{R}^{4}$ is, as for $m$-crystals, the set of dislocation invariants. 
     
     Because the continuous group elements can take small values, one expects that the energies of the corresponding continuous forms are small.
     
      \subsection{$m$-disclinations}\label{forms-matt}
\subsubsection{General formulae} $H_m \sim SO(3)$ leaves the 4-momentum ${p}_m^a$ invariant and acts on a manifold ${P}_{m\bot}$ orthogonal to $p_m^a $, which is a 3-space since $p_m^a $ is timelike. Any direction $\varpi ^{a} \subset {P}_{m\bot}$ is an axis of symmetry for continuous rotations. We choose $p_m^a $ as the time coordinate axis, $p_m^a = \{p^0_m,0,0,0 \}$, without loss of generality. Then $A_m=\begin{array}{|cc|} a     &  b  \\  -\bar{b}    &   \bar{a} \end{array}\,\,\mathrm{with}\, \|a\|^2+\|b\|^2=1$ is an element of $SU(2)=\widetilde{SO(3)}$ (cf. Sect.~\ref{PGr}).

Denote $x_{m}^a=\{x_m^0,x_m^1,x_m^2,x_m^3 \}$ any point on the cut manifold $\Sigma_m$, whose corresponding matrix representation reads ${x_m}=\begin{array}{|cc|}
 x_m^0+x_m^3   &\,\,    \bar{n} \\
   {n} & \,  x_m^0-x_m^3 
\end{array}$. $A_m$ rotates all together any point $\textbf{x}_m \in \Sigma_m ,\, \textbf x_m \rightarrow \textbf{x}'_m$ and the  causal structure attached to $\textbf{x}_m$ about the 4-momentum ${p}^a_m$ . Using Eq.~\ref{e2}, ($x_{m}'= A_m \, x_{m} \,A_m^*$), one gets:
\begin{multline}
\label{e7}
x_{m}'= 
 \begin{array}{|cc|}
(1 -2b \bar b)\,x_m^3+x_m^0 +(\bar a  b\, n+c.c.)  &\,   a^2\,{\bar n}-b^2 n - 2ab \,x_m^3 \\
\overline{a^2}\,{{n}}-\overline{b^2\,n} - 2\overline{ab}\, x_m^3 & \,\,\,   (-1+2b \bar b)\,x_m^3 +x_m^0 -(\bar a  b\, n+c.c.) \end{array}\\=\begin{array}{|cc|}
x_m^3 \,\cos 2\psi +x_m^0 +\rho \,\sin 2 \psi \, \cos u  &\,   e^{-i\omega}\left[\rho \,( \cos 2 \psi\, \cos u - i \sin u)-x_m^3\,\sin 2 \psi \right] \\
e^{i\omega}\left[\rho \,( \cos 2 \psi\, \cos u + i \sin u)-x_m^3\,\sin 2 \psi \right] & \,\,\,   -x_m^3 \,\cos 2\psi  +x_m^0 -\rho \,\sin 2 \psi \, \cos u
\end{array}.
\end{multline}
 \indent According to Eq.~\ref{e7}, $x_m'\mspace{0.25mu}^0=x_m^0$, so that $x_m^0$ can be taken as the cosmic time $\tau = x_m^0$ of the 3-space ${P}_{m \bot}$. We finally have
$${x'}_m^{1}+i{x'}_m^{2}=e^{i \omega}\left[\rho\, (\cos 2\psi\, \cos u +i \sin u)-x_m^3\,\sin 2 \psi \right],$$
\begin{equation}
\label{e8}
{x'}_m^0=x_m^0,\quad{x'}_m^{3}=\rho\,\sin 2\psi\,\cos u + \,x_m^3 \cos 2 \psi,
\end{equation}
where
$ a=\cos \psi\, e^{-i\frac{\alpha}{2}},\, b=\sin \psi\,e^{-i\frac{\beta}{2}}, \, n=x_m^1+i\,x_m^2=\rho e^ {i\theta},  \,u= \frac{\alpha -\beta}{2}+\theta,\, \omega =\frac{\alpha +\beta}{2}.$ 

 \indent Notice that $d_m=x'_m - x_m$ is always a spacelike displacement. We have indeed $d^0_m={x'}_m^0-x_m^0=0$, so that $\|{{d}^a_m}\|^2=(d^1_m)^2+(d^2_m)^2+(d^3_m)^2> 0$. In fact, all the displacements take place in a 3-space orthogonal to $p_m^a $, and result from the action of the $SO(3)$ group elements. Therefore one expects that the analysis of the $m$-disclinations does not differ from the analysis of continuous disclinations in a Euclidean 3-space, i.e., from the analysis of the disclinations in an amorphous medium. An example how the type of analysis made in $E^3$ can inspire the spacetime case is now given.
 
  \small{\textit{Notation}: from this point we denote $\lambda^{(i)}$ a i-dimensional form (e.g. $\lambda^{(1)}$ is a line-disclination or a line-dislocation, $\lambda^{(2)}$ a wall-disclination or a wall-dislocation, $\sigma$ a 2D cut surface bordered by a $\lambda^{(1)}$ and $\Sigma$ a 3D cut manifold bordered by a $\lambda^{(2)}$. }

\subsubsection{The wedge disclination $a\neq 0,\,b=0$}\label{$m$-disclinationlines1}
To get a physical insight of what a $m$-disclination is, we consider the limit case $\psi=0,\,a=\exp- i\alpha/2,\,b=0$.

 This is obviously a rotation of  angle $\alpha$ (possibly infinitesimal) about the $\{\textbf e_0\,\textbf e_3\}$ 2-plane. If $\{x_m^0,{x}_m^1,x_m^2,x_m ^3\}$ is an event on the cut manifold, its displacements 'upward' by an  angle $+\alpha /2$ and 'downward' by an  angle $-\alpha /2$ read:
$$({x}^\pm_m)^a =\{x_m^0,x_m^1\,\cos \frac{\alpha}{2}\mp x_m^2\,\sin \frac{\alpha}{2},\pm x_m^1 \,\sin \frac{\alpha}{2}+x_m^2 \,\cos \frac{\alpha}{2},x_m^3\},$$
\begin{equation}
\label{e9}
d^a_m=({x}^+_m)^a-({x}^-_m)^a=\{0,-2\,x_m^2\,\sin\frac{\alpha}{2} ,2\,x_m^1\,\sin\frac{\alpha}{2} ,0\}.
\end{equation}
\begin{figure}[h]
\includegraphics[width=6.5 in]{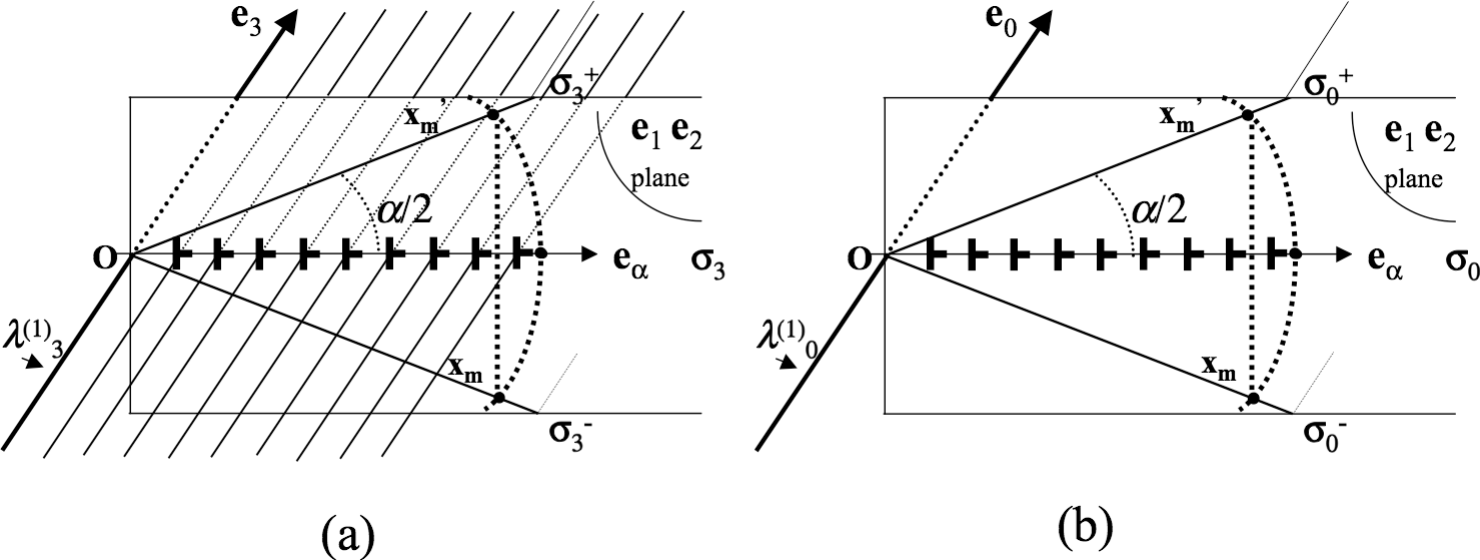}
\caption{Wedge disclinations $\lambda ^{(1)}_3$ Fig.~\ref{fig04.pdf}(a) and $\lambda ^{(1)}_0$ Fig.~\ref{fig04.pdf}(b), and their edge dislocation contents, schematized by the usual half cross symbol, drawn in the cut surfaces $\sigma_3$ and $\sigma_0$.  The dislocation lines are indicated in (a), but not in (b). $|\mathbf{ox}_m|=|\mathbf{ox}_m^{-}|=|\mathbf{ox}_m^{+}|$; $\textbf{d}=\textbf{x}_m^{-}\textbf{x}_m^{+};\,\,|\textbf{x}_m^{-}\textbf{x}_m^{+}|=2\, |\mathbf{ox}_m|\,\sin\dfrac{\alpha}{2}.$ The cut surface $\sigma_3$ (resp. $\sigma_0$) of the disclination in ${P}_{m\bot}$ (resp. in $\{ \textbf e_0\,\textbf e_1\,\textbf e_2 \}$) is along the $\{ \textbf{e}_\alpha\,\textbf{e}_3 \}$ plane (resp. the $\{ \textbf{e}_\alpha\,\textbf{e}_0 \}$ plane), and is bordered by the ${\lambda ^{(1)}_3}$ wedge line (resp. the ${\lambda ^{(1)}_0}$ wedge line).  The 3D cut manifold $\Sigma_m$ is the half timelike manifold $\{\textbf e_0\,\textbf e_3\,\textbf e_\alpha \}$ bordered by the  $\{ \textbf e_0\,\textbf e_3 \}$ plane, the $\lambda ^{(2)}$ wall disclination.}
 \label{fig04.pdf}
\end{figure}
\indent We notice that $d^a_m$ is a constant displacement vector on a $\{ \textbf{e}_0\,\textbf{e}_3 \}$ plane intersecting the $ \{ \textbf{e}_1\,\textbf{e}_2 \}$ plane at the point $\{{x}^1_m ={d^2_m}/({2 \sin \frac{\alpha}{2}}),\,{x}^2_m=-{d^1_m}/({2\sin \frac{\alpha}{2}})\}$. This can be interpreted as follows:

Since $d^a_m \in {P}_{m\bot} (\tau)$, consider in Eq.~\ref{e9} only the components that belong to ${P}_{m\bot} (\tau)$. Then Eq.~\ref{e9} reads:
 \begin{equation}
\label{e10}
 \textbf{x}^+_m -\textbf{x}^-_m =2\sin \frac{\alpha}{2}\,\textbf{e}_3 \times \textbf{x}_m,\quad \textbf{x}_m=\{{x}^1_m,{x}^2_m,{x}^3_m\}.
\end{equation}
      Thus the $\{ \textbf e_0\,\textbf e_3 \}$ axial plane of rotation for the $\{ \textbf e_1\,\textbf e_2 \}$ plane, when projected onto ${P}_{m\bot}(\tau)$, yields a rotation about the $\textbf{e}_3$ axis; it also yields a rotation about the $\textbf{e}_0$ axis when projected onto the 3 timelike submanifold $\{\textbf e_0\,\textbf e_1\,\textbf e_2\}$. This is represented Fig.~\ref{fig04.pdf}, where ${\lambda ^{(1)}_3}$ stands for the intersection of ${\lambda ^{(2)}}$ with ${P}_{m\bot}$ Fig.~\ref{fig04.pdf}(a), and ${\lambda ^{(1)}_0}$ stands for the intersection of ${\lambda ^{(2)}}$ with $\{\textbf e_0\,\textbf e_1\,\textbf e_2\}$ Fig.~\ref{fig04.pdf}(b).
                
    Consider a cut surface $\sigma_m \subset {P}_{m\bot}(\tau =x_m^0)$, (temporarily denoted $\sigma_3$ in Fig.~\ref{fig04.pdf}(a)) which extends away from the $\textbf e_3$ axis along a direction containing the point $\{ x_m^1,\,x_m^2\}$; the wedge opened (or closed, according to the sign of $\alpha$) is filled with (or deprived of) matter, as the Volterra process for a disclination requires, but this operation can also be achieved by a set of infinitesimal edge dislocations parallel to the axis of rotation $\textbf e_3$, whose Burgers vector total between the line and a line passing through $\textbf x _m$ is  $ \textbf{b}_m = \textbf x _m^+ -\textbf x _m^-$, with $d\textbf{b}_m/d{x}^\alpha = 2\sin \frac{\alpha}{2}\,\textbf{e}_3 \times \textbf{e}_\alpha$, i.e., equal to the displacement introduced by the disclination. This is sketched Fig.~\ref{fig04.pdf}(a). Similar considerations hold for the wedge disclination about the $\textbf e_0$ axis, Fig.~\ref{fig04.pdf}(b). 
    
  The wedge region in Fig.~\ref{fig04.pdf}(a) has to be understood as a region of accretion or of depletion $-$ according to the sign of the disclination $-$ of massive particles. %with a constant density. 
  The wedge region in Fig.~\ref{fig04.pdf}(b) is similarly filled by (or deprived of) a set of timelike geodesics carrying the particle 4-momenta $p_m^a$ parallel to $\textbf e_0$. The mass content of the wedge can be continuously modified by the arrival or departure of continuous $m$-dislocations moving along the $\{\textbf e_3 \,\textbf{e}_\alpha \}$ plane and the $\{ \textbf e_0 \,\textbf{e}_\alpha \}$ plane. But the 4-momenta $p_m =\{1,0,0,0\}$ do not suffer any rotation. The same conclusion holds for a general $m$-disclination.
  
  \textsl{Remark} The equivalence between a wedge disclination and a set of edge dislocations (a \textsf{tilt boundary}) is topological, which means that the overall geometry can be achieved either by the presence of a wedge disclination, or by a set of dislocations, or a mixture of both, insofar as both types of Volterra defects are allowed by the symmetries. But a caveat is here necessary. The dislocations are real defects, which mark the cut surface on which they are installed. Furthermore, as already alluded to, dislocations carry torsion \cite{hehl}, whereas disclinations carry only curvature. Insofar as a disclination is topologically allowed, one expects that it is achieved without dislocations.

\subsubsection{{$m$- grain boundaries as imperfect $m$-disclinations}}\label{m-disclinationlines3}
The misorientation introduced by a $m$-disclination from one side of the cut manifold $\Sigma_m$ to the other is obtained by comparing $p$ on one side to $p' = A_m \,p\, A_m^*$ on the other. For example $p_r^a=\{1,1,0,0\} \rightarrow p'_r=A_m \,p_r\, A_m^* =\{1,\cos \alpha,\sin \alpha,0\}$, with $A_m$ as above with $\psi=0$. $A_m$ does not leave $p_r^a$ invariant. The $m$-disclination characterized by the group element $A_m \in \widetilde{SO(3)}$ and the cut manifold $\Sigma_m$ does not satisfy the Volterra symmetry requirements for the $r$-crystal defined by $p_r^a$.

In fact, the full causal structure is rotated rigidly from one side of $\Sigma_m$ to the other, about the invariant direction of the 4-momentum $p_m =\{1,0,0,0\}$. The consideration of such a misorientation makes sense if a $p$-congruence is physically occupied by the corresponding substance. In such a case the cut manifold appears as a region of rapid variation of the  $p$-crystal, i.e. as a wall in 3-space. 

The term $m$- \textsf{grain (sub)boundary} is borrowed from CMP, where a grain boundary is a portion of wall separating two crystal grains 
misoriented one with respect to the other; 'sub' relates to the fact that the misorientation may be small. The border of this wall can be considered as an \textsf{imperfect $m$-disclination}, since it does not satisfy the Volterra symmetry requirements for the $p$-crystal. 

An imperfect $m$-disclination can be defined by a set of imperfect constitutive dislocations that fill its cut manifold, in the manner advanced for a wedge disclination in the foregoing paragraph \cite{klfr08}. We do not dwell on that description, which is equivalent to the description given above, without bringing anything essential (the same is true for imperfect $r$-disclinations).

 \subsection{$r$-disclinations}\label{forms-light}
\subsubsection{General formulae}The general expression for an element $A_r$ of $\widetilde{H_r}$ that conserves the null 4-momentum ${p}^a_r=\{p^0_r,0,0,p^3_r\},\,\,p^3_r=p^0_r$ can be written \cite{wigner39%,sternberg94
}:
  \begin{equation}
\label{e11}
A_r=\begin{array}{|cc|}
  \exp{-i\frac{\alpha}{2}}   \, \,\,&\,  i\bar{\omega}\,\exp{i\frac{\alpha}{2}} \\
     0 \,\,&\,   \exp{i\frac{\alpha}{2}}  
\end{array},
\end{equation}  
where $\omega=\omega_1+i\omega_2$ is a complex representation for a 2D vector
 which can take any value, and $0\leqslant{\alpha} <4 \pi$. $A_r$ rotates all together the point $\textbf{x}_r=\{x_r^0,x_r^1,x_r^2,x_r^3\}$ of the cut manifold $\Sigma$ and the  causal structure attached to $\textbf{x}_r$ about the null 4-momentum ${p}^a_r$. 
 
 The displacement of the cut manifold reads, employing Eq. \ref{e2}:
 \begin{equation}
\label{e12}
  d_r =x'_r -x_r= 
 \begin{array}{|cc|}
(e^{i {\alpha}}i\bar \omega\,{n}+c.c.)+\omega \bar{\omega}\,(x_r^0-x_r^3)  &\,\,\,\,\,\,   \bar{n}(e^{- i\alpha}-1)+i\bar{\omega}(x_r^0-x_r^3) \\
{n}(e^{ i\alpha}-1)-i{\omega}(x_r^0-x_r^3) & \,\,\, \,\,\,\,\,\, 0
\end{array},
\end{equation} 
where $n=x_r^1+i x_r^2.$

Since $d_r^0=d_r^3$, $d_r$ is spacelike, as $d_m$.

The $r$-disclination describes a region of concentration (a source) or of deficit (a sink) of radiation.
The direction of $p_r$ is invariant, but any other direction is rotated. We do not elaborate as the situation is comparable to the $m$-case.

\subsubsection{{$r$- grain boundaries as imperfect $r$-disclinations}}\label{r-disclinationlines3}
As above in Sect.~\ref{m-disclinationlines3}, the misorientation introduced by a $r$-disclination from one side of the cut manifold $\Sigma_r$ to the other is obtained by comparing $p$ on one side to $p' = A_r \,p\, A_r^*$ on the other. There is of course no misorientation for the 4-momentum $p_r =\{1,0,0,1\}$, but any other momentum $p$, whatever its nature (timelike, null, or spacelike) is tilted, when crossing $\Sigma_r$. Again, the consideration of such a misorientation makes sense if the $p$-congruence is occupied by the relevant substance, making it a $p$-crystal. The cut manifold appears as a region of rapid variation of $p$, i.e. a wall in 3-space. The border of this wall can be considered, in analogy with Sect.\ref{m-disclinationlines3}, as an imperfect $r$-disclination, and the cut manifold as a $r$- grain boundary. 

 \subsection{Dislocations} \label{dislocationforms}
 
 \subsubsection{About the nature of Burgers vectors and dislocation cut manifolds}   
   The effect of a dislocation form introduced in $M^4$ depends on two factors, its \textsf{Burgers vector} $b^a$, and the shape of its cut manifold $\Sigma$. At any point $ \textbf x \in \lambda^{(2)}=\partial \Sigma$ we split $b^a$ into its projection $\kappa^a$ on the tangent plane of $\partial \Sigma$ at $\textbf x$ and a vector $\mu^a$ orthogonal to $\kappa^a$, $b^a=\kappa^a+\mu^a, \, \kappa^a\,\mu_a=0$.

(a)-  \textit{$\kappa^{a}$ timelike}.  Assume that $\kappa^{a}$ is timelike, which we denote $ \kappa^{a}=t^a$, and joins two points $\textbf x,\,\textbf y \in \lambda^{(2)}$. We claim that this is a unrealistic physical situation.  Indeed there exists an open neighborhood V $\subset M^4$ including $\textbf x$ and $\textbf y$, which contains a continuous set of spacelike 3-surfaces $\Pi_i$ orthogonal to a timelike geodesic segment $\in$ V containing $\textbf x \textbf y$. The $\Pi_i$'s intersect $\Sigma$ along 2D spaces $\sigma_i$'s orthogonal to $t^a$, which are cut surfaces in the $\Pi_i$'s. Locally, the Volterra process is geometrically equivalent to a relative slip by $t^a$ of the two parts of the spacelike 3-surfaces with respect to  the intersections $\sigma_i$. Therefore the Volterra process transforms a set of parallel 3-space hypersurfaces $\subset$ V, at a repeat distance $\|t^a\|$, into a unique hypersurface having the shape of an helicoid. This is reminiscent of the topology of a screw dislocation in CMP \cite{friedeldisloc}. 

But while screw dislocations are very common objects in condensed matter, one might suspect that they do not occur in GR if the displacement vector has a timelike component, because the presence of a spacelike helicoid is in conflict with the notion of a well defined cosmic time. \textit{Thus the projection $\kappa^a$ of the Burgers vector on a wall-dislocation could not be timelike.}
    
 (b)- \textit{Achronal cut manifolds.} In order to avoid such a blemish, we restrict our consideration of wall-dislocations $ \lambda^{(2)}$ with $t^a \neq 0$ to those $\lambda^{(2)}$'s that do not contain timelike directions; they can always be considered as boundaries of \textsl{achronal} cut manifolds. A 3D manifold $\Sigma \subset M,\, M \,\mathrm{a \,spacetime,}$ is {achronal} if no two points $\in \Sigma$ can be joined by a timelike trajectory. Achronal hypersurfaces are of importance in the HP theory of singularities.
 
 An important notion related to achronal surfaces is that one of \textsl{domain of dependence} \cite{geroch70}. It is the set of events $p$ that are entirely determined by $\Sigma$, considered for the present purpose as the cut manifold of a line defect in GR, 
 and made of  $D^+(\Sigma)$, the set of future events depending causally on all the events belonging to $\Sigma$, and  $D^+(\Sigma)$ (past events).
  The boundaries $H^+(\Sigma)$ of $D^+(\Sigma)$, and $H^-(\Sigma)$ of $D^-(\Sigma)$, $\Sigma$ excluded, are achronal null 3-surfaces generated by
   the null geodesics that end on $\partial \Sigma$. $H(\Sigma)=H^+(\Sigma)\cup H^-(\Sigma)$ is the Cauchy horizon of $\Sigma$. \textit{Before} (in the past), $\Sigma$ represents the experimentator choice of a location for the $m$-form $\partial \Sigma$ in $M^4$, say; \textit{After} (in the future), $\Sigma$ might be taken as the mid achronal surface between the achronal sets $\Sigma^+$ and  $\Sigma^-$. 
   Any event $p$ inside the domains of dependence can be joined to $\Sigma$ by a timelike geodesic that intersects $\Sigma$, but not to $\partial \Sigma$, now transformed to a singular region of the spacetime, better represented by a cavity whose topology $\partial \Sigma \times S^1$ has been described above. 
   
   The null trajectories that belong to the Cauchy horizon of $\Sigma$ and that reach $\partial \Sigma$ are incomplete \cite{wald84}. They are analogous, in terms of defect theory, to the incomplete rows of atoms alluded to above, Fig.~\ref{fig03.pdf}.}
 
 (c)- \textit{$\kappa^{a}$ spacelike.} If $\kappa^{a}=s^a$ is spacelike, then any $\Sigma \subset M^4$ is either achronal spacelike or,  if $\Sigma$ non-achronal, is defined locally by one timelike direction and two orthogonal spacelike directions; in that latter case $s^a$ belongs to the 2-plane spanned by these spacelike directions.
 
 (d)- \textit{$\kappa^{a}$ null.} There is a family of null trajectories $\subset \Sigma $ and $\Sigma $ is achronal null. \\
 
 Notice that the above considerations can be extended to the displacements $d^{a}=b^{a}+L(\mathbf x) : \mathbf x$. We have seen above that the displacements $d^{a}_m$ and $d^{a}_r$ are both spacelike, so that there is no limitation on the choice of disclination cut manifolds.
 
\subsubsection{$m$-dislocations; interplay with $m$-disclinations}\label{mdislo} The situation is akin to that one met for a glass in Riemannian 3D manifold of constant curvature. According to the analysis carried in ref. \cite{klfr08}, the Volterra disclinations belonging to such a space curve under the action of continuous Volterra dislocations attached to them, whose Burgers vectors belong to the same space. In the present case in $M_4$, we have shown above the existence of disclinations belonging to a 3-space submanifold which carries an invariant rotation belonging to $SO(3)$; this group transforms this subspace into itself, so that the attached dislocations belong to the same subspace. In other words the flexibility and mobility of the $m$-disclinations in $M_4$ are attended by the creation and annihilation of $m$-dislocations with a \textsl{spacelike} Burgers vector; they carry torsion, i.e. {spin angular momentum}. 

\subsubsection{$r$-dislocations} \label{rdislo}They are not topologically distinct from $m$-dislocations, insofar as we focus our interest on the changes in the spacetime metric, i.e. on the effect of the presence of the dislocation on the variation of curvature (mass) and torsion (angular momentum). Again, in order that the constant cosmic time surfaces not be transformed into helicoids, it is necessary that the cut manifold does not contain timelike Burgers vector components, which requires $\Sigma$ to be achronal. Also, one can remark, as in the case of $m$-forms, that the displacement vector of $r$-disclinations being spacelike, the $r$-dislocations with which they interact have also spacelike Burgers vectors. 

\subsection{$h$-disclinations} \label{hypdiscli}  
   A \textsf{Lorentz boost} is the analog of a rotation; instead of the standard circular angle $\alpha$, we have an hyperbolic angle $\phi$. An hyperbolic rotation in the $\{\textbf e^0\,\textbf e^3\}$ plane reads
 \begin{equation}
\label{e13}
{x'}^0= x^0 \,\cosh \phi+{x}^3 \,\sinh \phi , \quad {x'}^3= x^0 \,\sinh \phi+ {x}^3 \, \cosh \phi , \quad {x'}^1=x^1, \quad {x'}^2=x^2. \end{equation} With $\cosh \phi = \gamma = (1-v^2)^{-1/2}$ one recovers the usual equations for the change of coordinates from one inertial frame to another with a velocity $\{0,0,v\}$ with respect to the first one.  Eq.~\ref{e13} can also be written as: 
\begin{equation}
\label{e14}  x' = A_h\, x\, A_h^*
,  \qquad \mathrm{with}\qquad
A_h=\begin{array}{|cc|}
 e^{\frac{\phi}{2}}     &  0  \\
 0    &    e^{-\frac{\phi}{2}} 
\end{array} =A_h^*.  \end{equation}

$A_h$ belongs to the Lorentz group $L_0$ (in fact to its universal cover $\widetilde{L_0}$) but not to the little groups $\widetilde{H_m}$ or $\widetilde{H_r}$; thus no \textsl{perfect} Volterra form can be associated to a boost in a $M_4$-crystal, in so far as we have rejected the possibility of tachyon $M_4$-crystals. Except that a boost can be split into a family of continuous $m$- or $r$-dislocations, which are allowed Volterra forms. In fact, while dislocations are just a useful trick in the case of wedge $m$- or $r$-disclinations\footnote{\label{f5}This claim must be qualified: infinitesimal $m$-dislocations with spacelike Burgers vectors play a distinctive role in all processes that involve movement, change of shape, relaxation, etc. of $m$-disclinations, in analogy with the role they play in CMP \cite{klfr08}. This was mentioned in Sect.~\ref{mdislo} and \ref{rdislo}. We do not expatiate on this subject.}, they are a \textsl{prerequisite} for the construction of boost (or hyperbolic) disclinations ($h$-disclinations).

Let $\{x_h^0,{x}_h^1,x_h^2,x_h ^3\}$ be an event on the cut manifold $\Sigma_h$; its displacement under the action of $A_h$ is:
\begin{equation}
\label{e15}
d^a=\big\{2\,\sinh\frac{\phi}{2}(x_h^0\,\sinh\frac{\phi}{2}+x_h^3\,\cosh\frac{\phi}{2}) ,0,0,2\,\sinh\frac{\phi}{2}(x_h^0\,\cosh\frac{\phi}{2}+x_h^3\,\sinh\frac{\phi}{2})\big\}.
\end{equation}
Because $\|{d^a}\|^2 = 4\,\sinh^2 \frac{\phi}{2} \, [(x_h^3)^2 - (x_h^0)^2 ]$, $d^a$ can be either spacelike, timelike or null, on the same cut manifold.

We haven't yet specified the $p_m$ and $p_r$ congruences that are affected by the boost disclinations. We noticed that for $m$- and $r$- disclinations (Sect.~\ref{m-disclinationlines3}, \ref{r-disclinationlines3}) there is only one 4-momentum that is invariant under the corresponding little groups $H_m$ and $H_r$; any other 4-momentum is misoriented at the cut manifold. Here there is an entire space 2-plane $\Pi$ ($\textbf e_1 \textbf e_2$ in the present case) which is invariant under a boost. Therefore the imperfect character of a $h$-disclination affects all the 4-momenta that belong to a timelike 2-plane $\Pi_\bot$ orthogonal to $\Pi$ ($\textbf e_0 \textbf e_3$ in the present case). 

\subsection{The form network and the large scale structure of the Universe} \label{network}
Forms can merge at nodes and thereby constitute a connected network. At each node Kirchhoff relations are obeyed, for example $\sum_a d^a = 0$ for suitably oriented dislocations; the Kirchhoff relations are more subtle for disclinations, cf. \cite{klfr08}. We shall not dwell on this question, but we notice that a connected lattice of perfect and imperfect disclinations presents a deep analogy with the filamentary and sheet-like large scale structure of the Universe; we surmise that the filaments originate as perfect disclinations and the sheets as imperfect ones.  One might fancy that another connected network originates as dislocations, but the empirical presence of sheets in the large-scale structure of the Universe, i.e., in the framework of the present interpretation, the presence of grain boundaries bordered by imperfect disclinations, favors an interpretation of the structure of the Universe in terms of disclinations. After all disclinations are Volterra defects attached to the $L_0$ part of the group of symmetry, which is more representative of any GR manifold than the translation part.

We therefore figure out the Universe as a single crystal with some mosaicity due to the misorientations at the grain boundaries. To see the essence of this mosaicity, consider a $r$-disclination whose radiative core has been transformed to a massive core at the electroweak transition. 
The cut manifold $\Sigma_r$ is now crossed by massive particles $p_m=\{p^0,p^1,p^2,p^3\};\,\|p_m^a\|^2 = -1$ which tilt from one side of $\Sigma_r$ to the other, since they see the $r$-disclination as an imperfect disclination. The 4-momentum $p_m$ is not much different from $p_r$, thereby $|p_m^0|$ and $|p_m^3|$ are large compared to $|p_m^1|$ and $|p_m^2|$. Furthermore we assume, motivated by the analysis of Sect.~\ref{dislocationforms}1, that $\Sigma_r$ is a null achronal surface, the radiation 4-momentum $p_r$ is then orthogonal to $\Sigma_r$. Thus it is tempting to identify $v^3=p_m^3/\gamma$ with Dirac's \textsf{natural velocity} (see Sect.~\ref{stronghyp}) and $v^1=p_m^1/\gamma$ and $v^2=p_m^2/\gamma$ with its observed orthogonal fluctuations, the so-called \textsf{peculiar velocities}. 

  According to Eq.~\ref{e12} (which is valid for the displacement of an event $\textbf x_r$ but also for the causal structure it carries) we have:
$$\triangle  p_m^{0}=\triangle  p_m^{3}=-\,|\omega| |p_{m\bot}| \, \sin(\phi - \phi_0 + \alpha)+ |\omega|^2 \, (p_m^0 =p_m^3) \approx  p^1_m \, \omega_2 -p^2_m \, \omega_1 =|\bm\omega  \times \textbf p_m|_3,$$ 
$$\triangle  p_m^{1}=-2\,\sin\frac{\alpha}{2} \, (p_m^1 \sin \frac{\alpha}{2} +p_m^2 \cos \frac{\alpha}{2}) \approx -\alpha \, p_m^2,\, \triangle  p_m^{2}=-2\,\sin\frac{\alpha}{2} \, (-p_m^1 \cos \frac{\alpha}{2} +p_m^2 \sin \frac{\alpha}{2})\approx \alpha \, p_m^1.$$ $\alpha$ and $\omega$ are small quantities and we have neglected second order terms; furthermore on expects that $p_m^{0}-p_m^{3}$ be small, since these 4-momentum components are the transforms of $p_r^{0}$ and $p_m^{3}$ at the electroweak transition .

$\triangle \textbf{p}=\gamma \triangle \textbf{v}$ measures the misorientation of the velocity in 3-space; the modulus ${\triangle v}=\gamma^{-1} \sqrt{(\triangle p^1)^2+(\triangle p^2)^2+(\triangle p^3)^2}$ also reads: 
    \begin{equation}
\label{e16}
{\triangle v}=\pm 2\,\sin \frac{\alpha}{2}\,\sqrt{v_1^2 + v_2^2+(|\bm\omega  \times \textbf p_m|_3)^2}.
\end{equation}

We interpret Eq.~\ref{e16} under the assumption that the strengths $\alpha,\, \bm \omega$ of the $r$-disclination is approximately the same all over the Universe, as an application of the cosmological principle; $\alpha$ measures the mosaicity. And indeed
these velocities seem to be approximately constant through the universe, with a mean square root $\textsc{V} =\sqrt{v_1^2 + v_2^2} \approx 340$ km/s (see \cite{wang-2007} and references therein).

Eq.~\ref{e16} has a remarkable feature: it is independant of the Dirac component of the velocity $v_3$. Identifying $\sqrt{v_1^2 + v_2^2+(|\bm\omega  \times \textbf p_m|_3)^2}$ with $\textsc{V}$ (assuming $\bm \omega$ small), we see that the misorientation $\triangle \textbf v$ and the modulus $\triangle v$ are approximately constant. 
In order to construct a detailed model of the large scale structure of the universe, relations between $\triangle v, \, p$, and strengths have to be established and discussed for the various types of perfect and imperfect forms, but this is 
a non trivial task that demands a critical examination of the relevant experimental data. 

\section{Discussion} \label{disc}
 
 We have tried in this article to analyze the defects (continuous or quantized) one might expect to find in the Universe; we call them \textit{forms}. Our analysis relies on the kinship between these forms on one hand, defects in the sense of Condensed Matter Physics on the other. In the same way that there are groups of invariance for each 'category' of ordered matter in the Euclidean space $E^3$ of CMP, we surmise the existence of groups of invariance in the spacetime (in this article we restrict to the Minkowski space) for the two categories of existing \textit{substances}, massive matter $m$ and massless radiation $r$.  These groups of invariance $H_m$ and $H_r$ are taken here as the symmetry groups of the congruences of geodesics of these substances, which is giving these congruences a physical reality not commonly considered. They are subgroups of the Poincar\'e group. Other more sophisticated groups of invariance might emerge from an analysis which would go beyond Wigner's classical investigation of the representations of the {Poincar\'e group} and their relationship with elementary particles. But we content ourselves with the present approach. 
  
    Furthermore, it is difficult to imagine a spacetime manifold, endowed with physical properties, without a substance. Here again, the comparison with CMP is rewarding. Consider an amorphous substance in $E^3$; its group of invariance is the full Euclidean group $E(3)$, and its defects can be classified by the homotopy groups of its coset space $-$ which in this case is trivial $-$, and the Volterra defects attached to $E(3)$, as we learn from a standard application of the CMP theory of defects. This statement makes sense because amorphous substances have a physical reality in $E^3$, which means that the physical space $E^3$ has the properties of uniformity and homogeneity of such an amorphous substance, and that the defects attached to $E^3$ are those of an amorphous substance. By analogy the existence of defects and singularities attached to the full Poincar\'e group $P(4)$ would require the existence of a substance that obey a cosmological principle extended to the full spacetime (i.e., is uniform and homogeneous in $M^4$). We know that such an extended cosmological principle does not hold empirically. Thus, that this generalized principle does not hold, and that no substance exists whose global structure is invariant under the Poincar\'e group, are \textsl{equivalent} claims.
      
  Our model, which assumes a physical reality to congruences (of matter, of radiation), thereby raises them to the dignity of 'crystals' in $M^4$, whose defects are precisely the forms alluded to previously. This is the central principle on which this work is founded. Its justification may be in the very existence of the Hubble expansion, which is interpreted here, in Dirac's line, as the evidence of massive particle quasi-constant 4-momenta in large enough cells, whose scale remains to be determined.
  
  Despite the analogies between defects in CMP and in General Relativity, there is seemingly a crucial difference: GR defects, that deform the ground state of the congruences, also affect the curvature and torsion of the habit spacetime. The fact that it is not so in CMP is due to the neglect of relativistic effects, which is an obviously justified assumption. \\
   
The principles of the present analysis of forms in $M^4$ being set, we find three categories of defects: 

(a)- continuous line defects in 3D (surface defects in 4D), whose displacement vectors are not quantized and can be infinitesimal; they are akin to the Volterra type dislocations and disclinations of CMP, i.e. their strengths are in one-one correspondence with the elements of the groups of invariance $H_r$ and $H_m$ of $r$- and $m$-substances. 
In the frame of the elementary analysis we have carried out, the nature of the Volterra cut manifolds $\Sigma_m$ and $\Sigma_r$ is an important physical ingredient; the discussion in Sect.~\ref{dislocationforms} suggests that $\Sigma_m$ and $\Sigma_r$ are generically different, viz. achronal spacelike for $m$-forms, achronal null for $r$-forms. A more complete study is all wanting. 

From a physical point of view, we advance that continuous forms provide geometric tools to describe the phenomena of accretion of matter and concentration of radiation (and their opposite). But while continuous accretion results in an instability (gravitational collapse), so that accreting forms are preferred to exhausting forms, sources and sinks of radiation are equally favored. 

\textsl{Remark}  The qualifier \textsl{continuous} in the expression \textsl{continuous line defects} applies to the group elements; however the defects themselves can be singular; this is not likely in    
CMD (the viscous spreading of the core increases the entropy of the system) but has to be seriously considered in GR for attracting massive particles, whose clustering is a factor of entropy production.

(b)- imperfect disclinations, that are continuous line defects in 3D but not Volterra defects. The cut surface on which the imperfect disclination line leans is akin to a CMP grain boundary.  Indeed they are analogous to 2D-walls of dislocations bordered by disclinations. We distinguish three types of imperfect disclinations;  two of them related to $m$- or $r$- invariance breakings, and a third one ($h$-disclinations) to Lorentz boosts, which do not correspond to any kind of substance in the Wigner classification of elementary particles.

(c)- topological quantized forms, which exhibit incomplete null geodesics as Hawking \& Penrose's singularities do (but we do not discuss further of a possible relationship). They are classified by the homotopy classes of the coset space $\mathcal{M}_r=P(4)/H_r$; again, the $m$-forms are trivial ($\forall i,\, \pi_i (\mathcal{M}_m)\sim\{1\}$). The topological $r$-forms are the following: (i)- \textsf{fleeting singularities}, i.e. singular {events} classified by the homotopy classes of $\pi_3 (\mathcal{M}_r) \sim \mathbb{Z}$, with corresponding different topologies of their accompanying \textit{configurations} (in the sense of \cite{michel}) in 3-space cuts;  (ii)- \textsf{point singularities} (in 3D), classified by the homotopy classes of $\pi_2 (\mathcal{M}_r)\sim \mathbb{Z}$.\\ 
  
 What follows is highly speculative. Aside from the conjectures just advanced on the cosmological significance of the quantized singularities, one might wonder whether the Universe inhomogeneities below the galactic supercluster scales are not in relation with the primordial existence of a network of defects connected at nodes, junctions, with a typical cell size $\approx 20$ Mpc, formed by disclination lines and walls. 
According to the Standard Model, several massless particles acquire mass at the electroweak transition.Therefore forms of radiation might evolve  into forms of matter, at the very epoch when this transition does occur. This transformation is all the more probable as some of the corresponding continuous $r$- and $m$-forms are {topologically equivalent}. 
 0ne can therefore wonder whether the present inhomogeneous state of the universe, with filaments and walls, is not a relic of the defect network of the  radiation hot bath.
 
 Of course, that remark opens the question of the creation of the forms of radiation in the primordial universe at the so-called quark epoch. Since hot photons are strongly interacting with the quark-gluon plasma but also between themselves, it is hard to believe that the above $r$-crystal model is valid at this epoch. But we have no indication which structural model they obey. 
   
\subsection*{Acknowledgments}I thank J. Friedel, T. Damour, R. Omnès, R. Parentani,  J.-P. Poirier and M. Veyssié for discussions and encouragements, and R. Kerner for stimulating remarks at an early stage of this work.
%\newpage
\vspace{30pt}
\bibliography{gr-qc/0905.4643v2}

  \end{document}